\documentclass[5p]{elsarticle}

\usepackage{graphicx}
\begin{document}

\title{Precision timing measurement of phototube pulses \\ using a flash analog-to-digital converter}
\author[iu]{J.~V.~Bennett}
\author[iu]{M.~Kornicer}
\author[iu]{M.~R.~Shepherd\corref{cor1}}
\ead{mashephe@indiana.edu}
\cortext[cor1]{Corresponding author}
\author[jlab]{M.~M.~Ito}
\address[iu]{Indiana University, Bloomington, Indiana 47405, USA }
\address[jlab]{Thomas Jefferson National Accelerator Facility, Newport News, Virginia 23606, USA} 
\date{\today}

\begin{abstract}
We present the timing characteristics of the flash ADC readout  
of the GlueX forward calorimeter, which depends on precise measurement of arrival time of
pulses from FEU 84-3 photomultiplier tubes to suppress backgrounds.  The tests presented were performed using 
two different 250 MHz prototype flash ADC devices, one with eight-bit and 
one with twelve-bit sampling depth.  
All measured time resolutions were better than 1 ns, independent of signal size, 
which is the design goal for the GlueX forward calorimeter.  For pulses with an amplitude of 100~mV the timing resolution is  $0.57\pm 0.18$~ns, while for 500~mV pulses it is $0.24\pm 0.08$~ns.
\end{abstract}


\maketitle

\section{Introduction}
\label{Intro}

The forward electromagnetic calorimeter (FCAL) of the GlueX experiment~\cite{gluex} at Jefferson Lab 
will use flash analog-to-digital converters (FADC) to detect signals from final-state photons, 
produced in photon-nucleon reactions.  The calorimeter will consist of 2800 $4~\mathrm{cm}\times4~\mathrm{cm}\times45~\mathrm{cm}$ type F8 lead glass blocks, each coupled to an FEU~84-3 photomultiplier tube (PMT).
Electrons and positrons produced in the electromagnetic shower from an incident photon generate Cherenkov radiation that will be converted to a current pulse by the PMT.
The PMT pulses will be digitized by twelve-bit 250 MHz multi-channel FADC boards~\cite{Barbosa07}.
Continuous digitization by the FADC preserves the signal pulse shape, 
enabling simultaneous measurement of total charge, which is the integral of the pulse, and signal arrival time.
The integrated charge is proportional to the energy that the incident particle deposited in the block
and the pulse shape can be used to determine the arrival time of the particle.

The pulse arrival time will be used to (1)~isolate final-state photons from out-of-time hits in the detector that 
result from the expected high level of electromagnetic background and (2)~determine the exact beam bunch that produced the collision of interest.  The electromagnetic background arises from the photon beam, 
which will be delivered at an average rate of $10^8~\gamma$/s incident on the target (at energies of approximately 9~GeV) with beam bunch spacing of 2~ns.  Predicted background rates on the FCAL modules closest to the beam axis are of order 1~MHz; elsewhere in the detector the background rate is at the order of 10~kHz.  Given the bunch spacing, a timing resolution of better than 1~ns for each FCAL block is desired to both suppress electromagnetic background and perform bunch identification.  Note that a typical event will illuminate multiple blocks, thereby providing numerous independent measurements of the event time, each with better than 1~ns resolution, which will allow a determination of the beam bunch that produced the event of interest.

In order to make optimal use of the data bandwidth available, it is critical that the timing measurement of the pulses be done in real-time as part of the digital signal processing of the FADC buffer for each event.
We will present two such processing algorithms to determine the timing information from a PMT signal,
both of which utilize samples from the leading edge of the PMT pulse. We demonstrate that both of these algorithms provide time resolution sufficient for the needs of the GlueX experiment.

\section{Hardware characteristics}
\label{Readout}

The readout of the GlueX forward calorimeter will consist of 
2800 Russian-made FEU~84-3 PMTs, each powered by a
24~V Cockroft-Walton base~\cite{Brunner}. The PMTs will be connected to 
multiple 16-channel 250 MHz twelve-bit FADC boards,
designed and produced by Jefferson Lab~\cite{Barbosa07}. 
For tests described in this article, we utilized a single-channel eight-bit 
250 MHz FADC developed at Indiana University in addition to a
prototype eight-channel twelve-bit 250 MHz FADC developed by Jefferson Lab.
We will briefly describe these two FADC devices and the pulse shape 
of the FEU~84-3 phototube below.

 \subsection{Single- and multi-channel FADC}
 \label{FADChardware}

The eight-bit single-channel FADC prototype has been implemented 
as a PCI card so that it can be tested using a standard 
personal computer. The FADC digitizer samples the input signal
at 4~ns intervals, and the samples are continuously written into 
a dual-port ring-buffer. 
This enables the extraction of the samples without interrupting the digitizer. 
The data are read out over the PCI bus. 
The three processes, digitization, extraction and read out,
are independent of each other and 
can happen simultaneously, which allows the device to have zero dead-time.
The analog input saturates for voltages above $\approx$1.15~V.

The twelve-bit 250~MHz FADC is based on the MAX1215 chip with a 1.45~V dynamic range.  Like the eight-bit board, this device is also a zero dead-time device.  It has the ability to read signals simultaneously 
from up to eight PMTs, with the digitization of all channels synchronized to the same clock.  The board is based on the VME64x platform.  More details on the operation and characteristics can be found in Ref.~\cite{Barbosa07}.

 \subsection{Characteristics of the FEU~84-3 PMT pulse}
 \label{PMTpulse}

\begin{figure}
\begin{center}
\includegraphics[width =\linewidth]{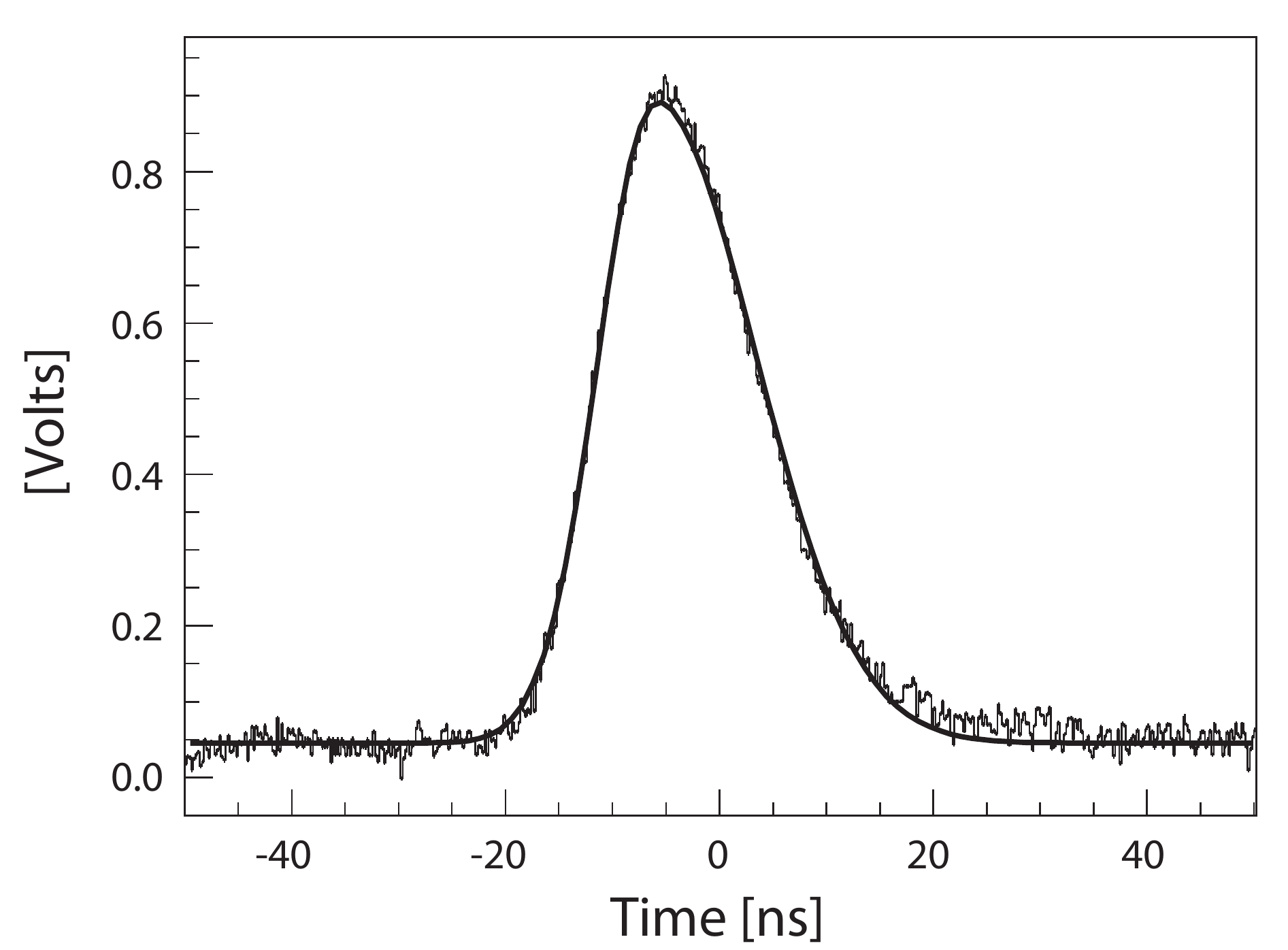}
\caption{ A typical pulse from FEU~84-3 PMT obtained from a digital oscilloscope. (The pulse has been inverted for comparison with FADC data.)  The solid line represents a fit to a bifurcated Gaussian with a constant offset.
}
\label{fig:pulse_shape}
\end{center}
\end{figure}

Figure~\ref{fig:pulse_shape} shows a typical pulse from a FEU 84-3 PMT 
obtained by digitizing the signal every 0.2~ns with a digital oscilloscope.
The line in Fig.~\ref{fig:pulse_shape} represents the fit 
obtained using a sum of a bifurcated Gaussian and a constant $B$, 
\begin{equation}
S(t)  = A \exp \left[ -\frac{\left ( t-t_p \right )^2}{2 \sigma_{k}^2} \right ] + B; ~~
k=
  \Big{\{} 
  \begin{array}{lr}
    1 & t < t_p  \\
    2 & t \ge t_p
  \end{array} ,
\label{eq:bigauss}
\end{equation}
where $A$ represents the amplitude of the signal, $t_p$ is the peak arrival time and 
$\sigma_{1,2}$ are the Gaussian widths that model the duration of the rising and falling edges, respectively.
It is evident that the rising edge is well reproduced by a Gaussian line-shape, and this
feature will be employed by one algorithm for the online extraction of the pulse arrival time. 
In addition, pulse characteristics are given by the parameters of the bifurcated Gaussian fit. 
We found that the characteristic leading-edge time ($\sigma_1$) is about 9~ns independent of the pulse amplitude.

Since the pulse varies slowly around the peak, the peak sample time is a poor choice to characterize the arrival time of the pulse.  We choose a characteristic signal time corresponding to the time at which the pulse has reached half of the signal maximum.  We call this time $t_{0}$ and discuss below two algorithms to measure $t_{0}$.

\section{Timing algorithms }\label{TimeAlg}

We explored two methods to obtain the characteristic time of a pulse
that both rely on two FADC samples on the rising edge of the pulse.  Other methods have
been used, {\it e.g.}, using the first algebraic moment~\cite{moment}; however,
we have had obtained the most precise results using the methods outlined below.  In the discussion
that follows we assume that all FADC measurements have been adjusted to remove the portion of the signal due to the DC offset, {\it i.e.}, all FADC samples are ``pedestal subtracted."

 \subsection{Gaussian transformation}\label{TwoSampleGauss}

This method takes advantage of the fact that the rising edge of the pulse from 
the PMT is Gaussian in shape.  Initially proposed by Teige {\it et al.}~\cite{Teige_time} it is based on the transformation that turns a Gaussian edge into a straight line:
\begin{equation}
S'_i = \sqrt{ -\ln \left ( \frac{S_i}{S_p} \right ) }, 
\label{eq:gauss_trans}
\end{equation}
where $S_p$ is the peak sample, while $S'_i$ is the transformed sample 
obtained from an original sample $S_i$ at the time $t_{i}$. 

The transformed rising edge samples are now a linear function of time
\begin{equation}
S'_i = a_{G} t_{i} + b_{G}.
\label{eq:SlinT}
\end{equation}
with the slope, $a_{G}$, and the intercept, $b_{G}$, related to the 
parameters of the rising-edge Gaussian, Eq.~\ref{eq:bigauss}.
The parameters of the straight line can be calculated using any two 
samples from the rising edge. The line, in turn, determines the time when the 
pulse reached any given fraction of its peak value.  In particular, $t_{0}$, is given by 
\begin{equation}
t_{0} = \frac{\sqrt{ - \ln{ \left( \frac{1}{2} \right ) } } - b_{G}}{a_{G}},
\label{eq:t50_def}
\end{equation}
It was found that the two samples immediately preceding the peak sample give
the best measure of $t_0$ in this approach~\cite{Teige_time}.

 \subsection{Linear interpolation}\label{TwoSampleLin}

It was desirable to test a different algorithm which did not involve mathematical operations such as 
a square-root and a logarithm in order to facilitate adaptation of the algorithm to the FADC field-programmable gate-array (FPGA).  For that reason we considered 
another method that assumes that the rising edge is a linear function of time in the vicinity of $t_0$.  A similar technique has been utilized with success by the VERITAS Collaboration~\cite{cogan}.
This assumption allows one 
to define the line by one sample preceding and the other one following half of the peak sample $S_p$.
These samples are labeled $S_-$ and $S_+$ respectively.
In this approach, the slope and intercept are given by
\begin{eqnarray}
a_L  &=&  \frac{S_+ - S_-}{ T }  \\ 
 b_L  &=&  S_+ - a_L t_+,
\label{eq:slope_lin}
\end{eqnarray}
respectively, where the time $t_+$ corresponds to the time of sample $S_+$ and $T$ is the sampling period, which is 4~ns for the two FADCs we studied.
The characteristic time is now determined by
\begin{equation}
t_{0}  = \frac{S_{p}/2 - b_{L}}{a_{L}}.
\label{eq:t50lin_def}
\end{equation}

In both approaches, we desire an algorithm that will provide a stable measurement of the characteristic pulse time independent of the pulse amplitude and other random variations in the pulse shape.  In all cases the timing information will be used comparatively, that is, we are interested in comparing the characteristic time of some channel to another channel in the detector array.  Therefore, {\em constant}, systematic offsets of $t_0$ from the true time at which the signal reaches half maximum are irrelevant.

\section{Timing resolution measurements}\label{TimeMeasure}

We tested the performance of the timing algorithms using an eight-bit single-channel and a twelve-bit multi-channel 
FADC.  With the single-channel FADC we examined digitization effects on the time resolution by effectively digitizing the same pulse twice with the FADC.  Such an approach removes statistical variations in pulse shape and explores the true resolution of the algorithm.  Using the multi-channel FADC we were able to recreate an environment comparable to that which we expect in the final detector.  By examining channel-to-channel differences in the algorithm-determined characteristic time of a common light source we were able to measure final timing resolutions for individual channels and explore the dependence of these resolutions on pulse amplitude.

 \subsection{Measurements with single-channel FADC}\label{Time8bit}

A single-channel eight-bit prototype was used to measure the timing resolution of the algorithms mentioned above.  
The light source in use was a small piece of scintillator, 
illuminated by a nitrogen laser LN300C.
The light from the scintillator was propagated via optical 
fiber into a dark-box containing an FEU~84-3 and an XP-2020 phototube. 
The operating voltage for FEU~84-3 was set at about 1600~V,
and the fast XP-2020 PMT was used as a trigger.

We devised a simple setup where the signal from a FEU~84-3 PMT was split, with one branch connected to the FADC and 
one end of the other branch left unterminated. The unterminated end produced a reflected pulse that was delayed, but still digitized within the same FADC buffer as the primary pulse.  If one neglects distortion and attenuation introduced by the delay cable, this technique produces two identical pulses with fixed separation in time and removes contributions to the resolution from fluctuations in shape or transit time in the PMT.  As long as the pulses are well separated in time, the algorithms described in Sec.~\ref{TimeAlg} can be applied to obtain 
the difference in characteristic time of two pulses $\Delta t_0$.  Since the true delay is fixed, the standard deviation of $\Delta t_0$ provides the quadrature-sum of the resolutions with which the characteristic times of the two pulses are determined.

\begin{figure}
\begin{center}
\includegraphics[width=\linewidth]{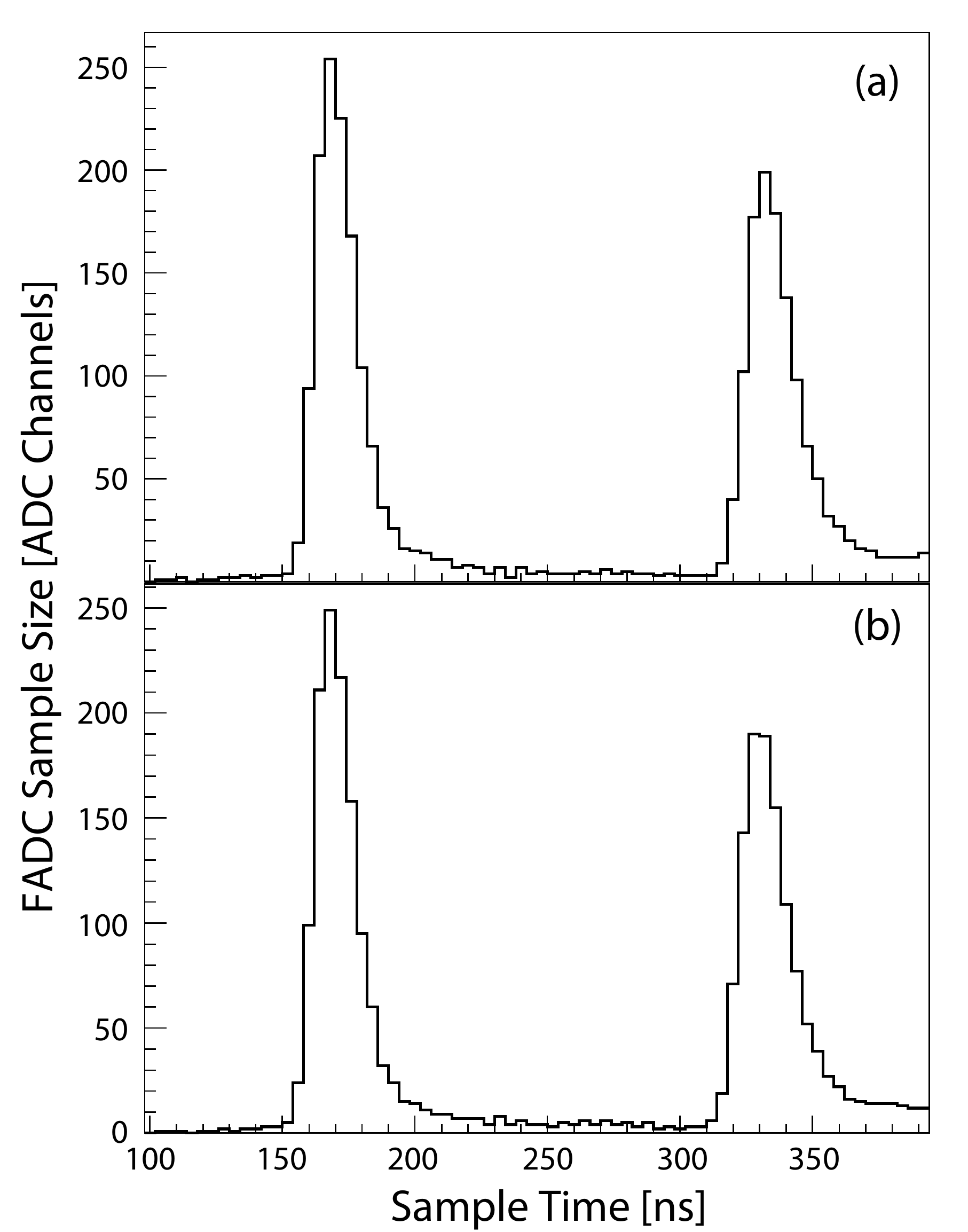}
\caption{ 
Typical FADC response for the primary and reflected pulse for two different delays.  Comparing the reflected pulse between panel (a) and (b) demonstrates what is termed in the text a difference in ``sampling phase."
}
\label{fig:twinPeak}
\end{center}
\end{figure}

Figure~\ref{fig:twinPeak} shows typical FADC responses for two different lengths of the open-end cable.  Comparing the reflected pulse between panels (a) and (b) of Figure~\ref{fig:twinPeak} demonstrates what we term a difference in ``sampling phase."  One can see that with Figure~\ref{fig:twinPeak}(b) the delay of the second pulse is such that the peak is likely to fluctuate between two adjacent samples, with the true peak being in the middle and slightly higher than the two.  As both algorithms depend on obtaining a sample at the peak, one anticipates some added uncertainty when the primary and reflected pulses are sampled with different phases.  By changing the length of the open end, one can control the delay between two pulses, modifying the probability that the sampling phase is different between the primary and reflected pulses, and explore the sensitivity of the timing resolution to the relative alignment of the samples with the structure of the pulse.  It is important to note that there is no synchronization between the FADC clock and the laser or trigger hardware, so it is impossible to control the absolute phase with which the pulses are sampled.

\begin{figure}
\begin{center}
\includegraphics[width=\linewidth]{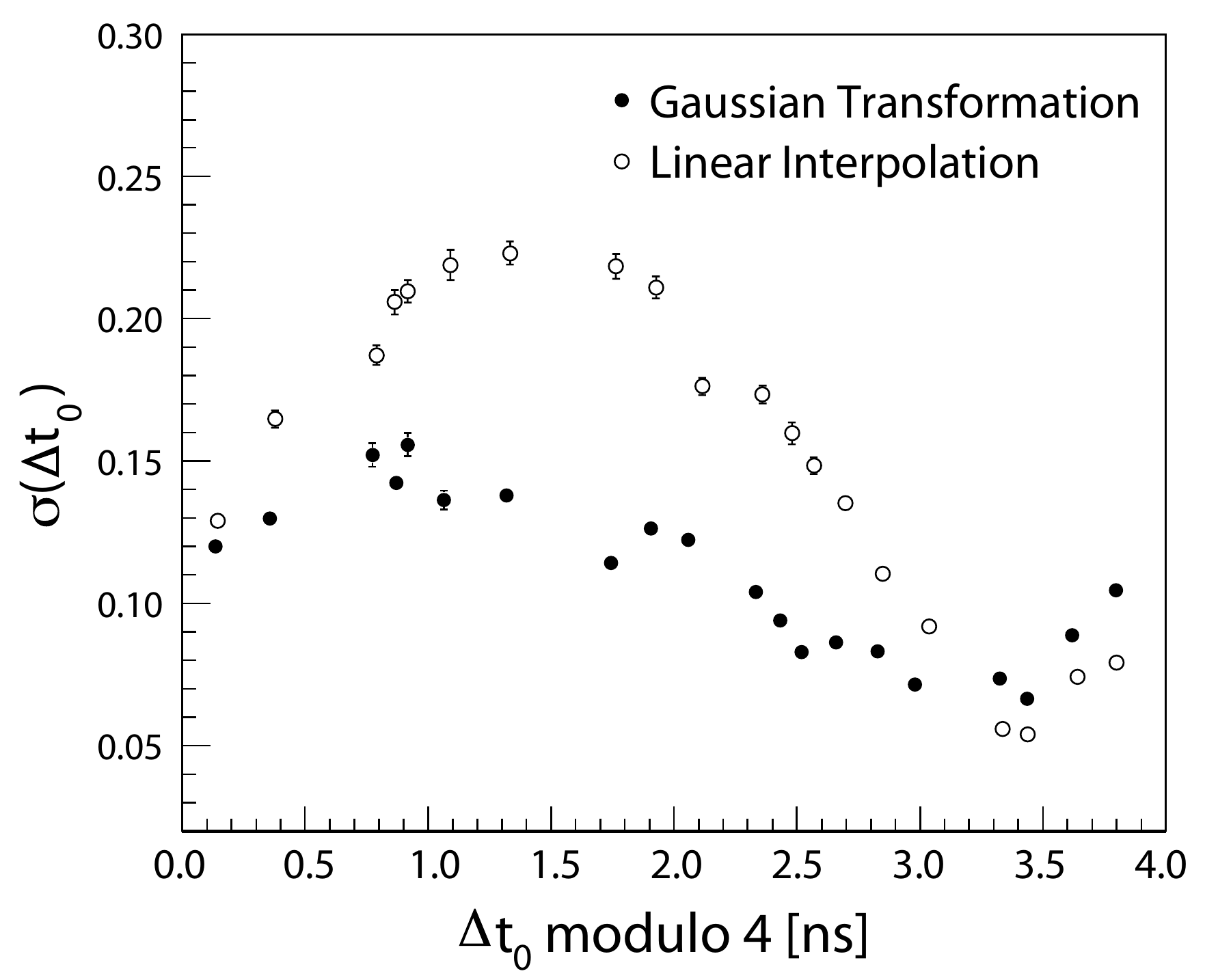}
\caption{\label{fig:twinDelay}
The value of $\sigma( \Delta t_{0})$ as a function of the mean delay ($\Delta t_{0}$) modulo 4 ns
obtained by applying Gaussian (filled circles) and linear-interpolation (open circles) methods.  The errors are statistical only.
}
\end{center}
\end{figure}

The dependence of the time resolution on length of the delay cable can be seen in 
Fig.~\ref{fig:twinDelay}, where, for various delays of the reflected pulse, we plot the standard deviation of $\Delta t_{0}$, $\sigma( \Delta t_0)$, obtained by applying the
Gaussian (filled circles) and linear-interpolation (open circles) methods.
The {\it x}-axis represents the mean of $\Delta t_0$ modulo 4 ns.  The best time resolution is obtained with a signal delay such that the peak region of both pulses are sampled with the same phase (as shown in Fig.~\ref{fig:twinPeak}(a)).  Naively, one would expect this to occur only when the delay cable length is a multiple of 4~ns; however, bifurcated gaussian fits to both the primary and reflected pulse indicate that the characteristic rise time, $\sigma_1$ in Eq.~\ref{eq:bigauss},  of the reflected pulse is approximately 0.5~ns larger than the first due to dispersion in extra length of cable.  This shifts the minimum of $\sigma(\Delta t_0)$ from 4.0~ns (or 0.0~ns) to about 3.5~ns.
The worst resolution from both methods occurs when the delay is changed by approximately 2~ns with respect to this minimum, corresponding to the case where both pulses are sampled with maximally different phases in the peak region (as shown in Fig.~\ref{fig:twinPeak}(b)).  In this case, the resolution in time-difference is larger for the linear-interpolation method by  about 70~ps. 

Assuming that $\sigma(\Delta t_0)$ arises from equal contributions of the timing resolution of the primary and reflected peak, we can estimate the time resolution for the algorithm itself
varies from 50 (40) to 110 (150) ps for the 
the Gaussian-transformation (linear-interpolation) methods. 

 \subsection{Measurements with multi-channel FADC}\label{Time12bit}

Pulse-to-pulse and channel-to-channel variations in transit time through the PMT and shape of the pulse will introduce fluctuations in the measured $t_0$ -- we call these fluctuations the ``statistical" contribution to the resolution as they arise from random effects.  We expect also an additional ``systematic" shift in the measured $t_0$ as a function of pulse height.  This systematic shift must be corrected in order to compare $t_0$ between two blocks in the array with different amounts of energy deposition.  The studies presented in the previous section only explore the resolution of the algorithm itself, hence a different approach is needed to examine these additional contributions to the timing resolution.

To better mimic the final detector configuration, we constructed an array of nine lead glass blocks optically coupled to PMTs which were powered by Cockcroft-Walton bases.  All of the blocks in the array were simultaneously illuminated by a pulsed LED, which was mounted in front of a diffusion panel oriented 
about 30 cm from the face of the lead glass array. The LED was driven by a pulsing circuit with a pulse 
rate of about 1~kHz.  A capacitive circuit was used to shorten the duration of the applied electrical pulse to the LED to less than 5 ns.  The entire setup was enclosed in a light tight box. 
One of the blocks in the experimental array was used as a trigger for the data acquisition system, 
while the remaining eight were connected to the FADC for data collection. 

Our goal was to extract the statistical and systematic contributions to the resolution as a function of pulse amplitude.
To reduce the pulse size, the incident light was attenuated.  In order to have a set of control data with which 
to compare those blocks with varying pulse amplitudes, a screen was placed in front of only four blocks with the 
other four uncovered.  The amplitude was varied by placing successively more screens in front of the 
appropriate blocks. The procedure was repeated using two different sets of control blocks.  In order maintain a relatively constant transit time through the PMT throughout the study, the bias voltage on the PMTs remained constant at the values in Tab.~\ref{tab:PMTvoltage}, which roughly equalized the gains for all PMTs.  (In the normal operating range of the FEU~84-3 PMT, variation in the bias voltage changes the transit time through the PMT by approximately 1~ns per 100~V change in bias voltage.)

\begin{table}
\begin{center}
\caption{PMT cathode voltage and pulse amplitude for various array elements.  The typical amplitude of the pulses was about 500~mV when no light-attenuating screens were installed.}
\label{tab:PMTvoltage}
\begin{tabular}{ccc}
\hline\hline
Channel & Voltage [kV] & Amplitude [Channels] \\
\hline
% add 4 to get back to Jake's original channel numbers
1 & -1.495 & 1376 \\
2 & -1.545 & 1398 \\
3 & -1.610 & 1414 \\
4 & -1.585 & 1405 \\
5 & -1.455 & 1374 \\
6 & -1.500 & 1401 \\
7 & -1.555 & 1371 \\
8 & -1.515 & 1439 \\
\hline\hline
\end{tabular}
\end{center}
\end{table}

\begin{figure}
\begin{center}
\includegraphics[scale=0.45]{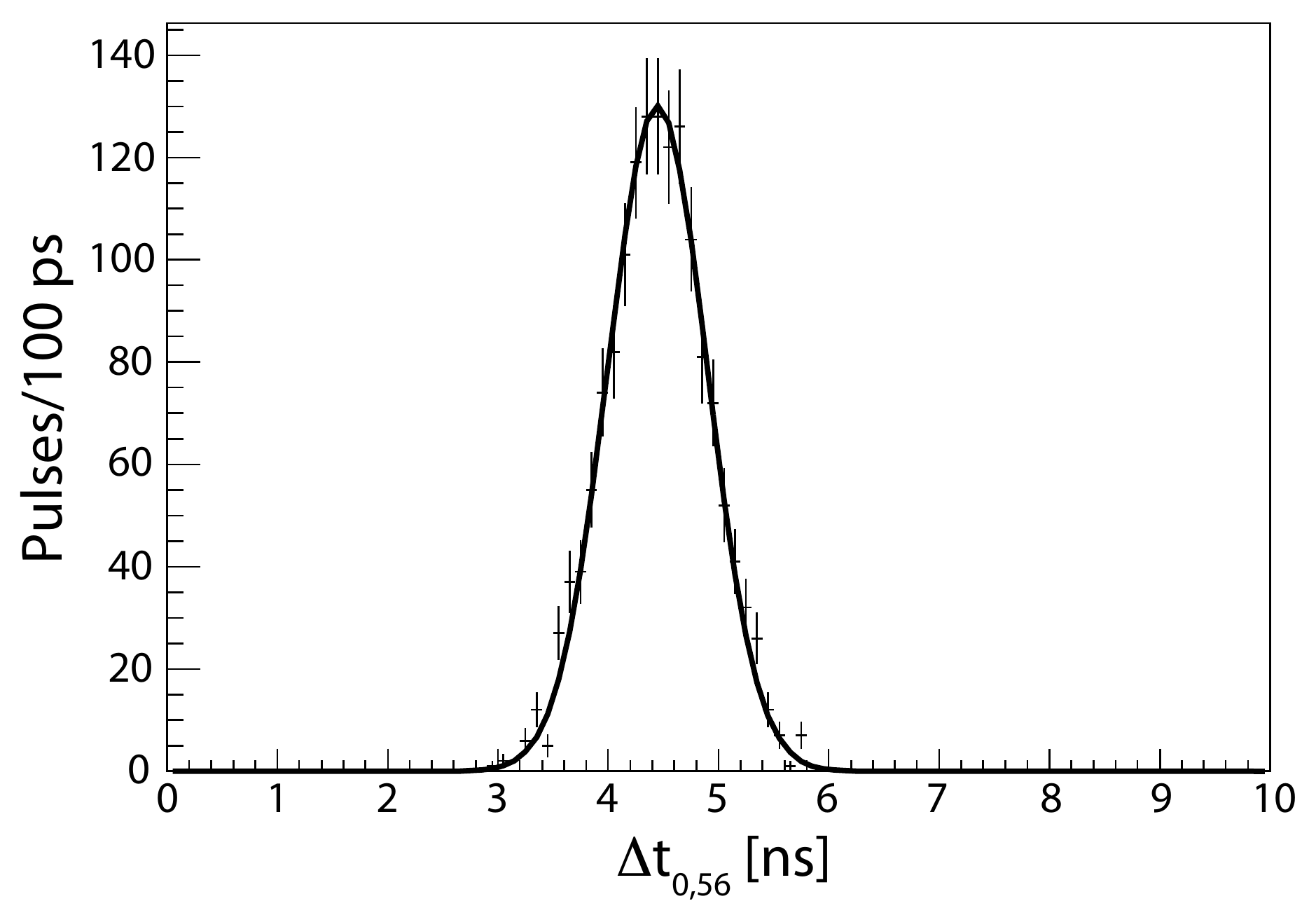}
\caption{This histogram of the difference in $t_{0}$ in two channels shows that the distribution is essentially
Gaussian in shape. Note that the two pulses do not arrive at exactly the same time.
\label{fig:dTGauss}
}
\end{center}
\end{figure}

As shown in Fig.~\ref{fig:dTGauss}, the difference in $t_{0}$ for any two channels $i$ and $j$ for a set of events 
can be characterized by a Gaussian distribution with mean $\Delta t_{0,ij}$ and standard deviation 
$\sigma_{ij}$. To obtain the mean value of the signal arrival time $t_{0,i}$ and statistical contribution to the resolution $\sigma_i$ for a single channel, a simple set of equations 
was used.  For $n$ independent channels there are, in principle, $n(n-1)/2$ unique, non-trivial equations of 
the form
\begin{eqnarray}
  \Delta t_{0,ij} &=& t_{0,i} - t_{0,j} \\
\label{eq:mu_ij}
  \sigma_{ij} &=& \sqrt{\sigma_i^2 + \sigma_j^2}. 
\label{eq:sigma_ij}
\end{eqnarray}
For a given configuration of the array, each pulse is digitized by the FADC and processed using the timing algorithms described in Sec.~\ref{TimeAlg}. 
The differences in $t_{0}$ for successive channels are fit to a Gaussian shape, from which 
 $\Delta t_{0,ij}$ and $\sigma_{ij}$ are extracted.  Collections of these fitted values are then utilized to study statistical fluctuations and systematic shifts in $t_0$ as a function of pulse height, as discussed below.

In order to obtain an accurate value of the statistical contribution to the resolution of a single channel 
as a function of signal size, we first determine the resolutions for the four control channels by using a $\chi^2$ fit.  For these four channels there are six independent values $\sigma_{ij}$. 
It is possible to choose a set of four $\sigma_i$ that minimizes a function given by
\begin{equation}
  \chi^2 = \sum_{i=1}^4 \sum_{j=i+1}^4 \left(\frac{\sigma_{ij} - \sqrt{\sigma_i^2 +
  \sigma_j^2}}{{\delta(\sigma_{ij})}}\right)^2, 
\label{eq:minChi2}
\end{equation}
where $\delta(\sigma_{ij})$ is the error in $\sigma_{ij}$ obtained from the Gaussian fit to the time-difference 
distribution. The minimization of this function yields values for the statistical contribution to the resolution.  Repeating for the other set of four channels gives resolutions for all eight blocks in the array (at pulse amplitudes of about 500~mV) as shown in Tab.~\ref{tab:sigma_vs_chan}.

\begin{table}
\begin{center}
\caption{Extracted $t_{0}$ standard deviations for the Gaussian transformation method
($\sigma_{G}$) and the linear interpolation method ($\sigma_{L}$) at pulse amplitudes of about 500~mV.  These measurements represent the dominant, statistical contribution to the resolution.}
\label{tab:sigma_vs_chan}
\begin{tabular}{ccc}
\hline\hline
Channel & $\sigma_G$ [ns] &  $\sigma_L$ [ns]  \\
\hline
%add 4 to get back to Jake's original channels
1 & $0.28 \pm 0.01$ & $0.25 \pm 0.01$ \\
2 & $0.31 \pm 0.01$ & $0.28 \pm 0.01$ \\ 
3 & $0.29 \pm 0.01$ & $0.25 \pm 0.01$ \\
4 & $0.30 \pm 0.01$ & $0.28 \pm 0.01$ \\
5 & $0.27 \pm 0.01$ & $0.26 \pm 0.01$ \\
6 & $0.30 \pm 0.01$ & $0.28 \pm 0.01$  \\
7 & $0.27 \pm 0.01$ & $0.26 \pm 0.01$ \\
8 & $0.29 \pm 0.01$ & $0.28 \pm 0.01$ \\
\hline\hline
\end{tabular}
\end{center}
\end{table}

\begin{center}
\begin{figure}
\includegraphics[width=\linewidth]{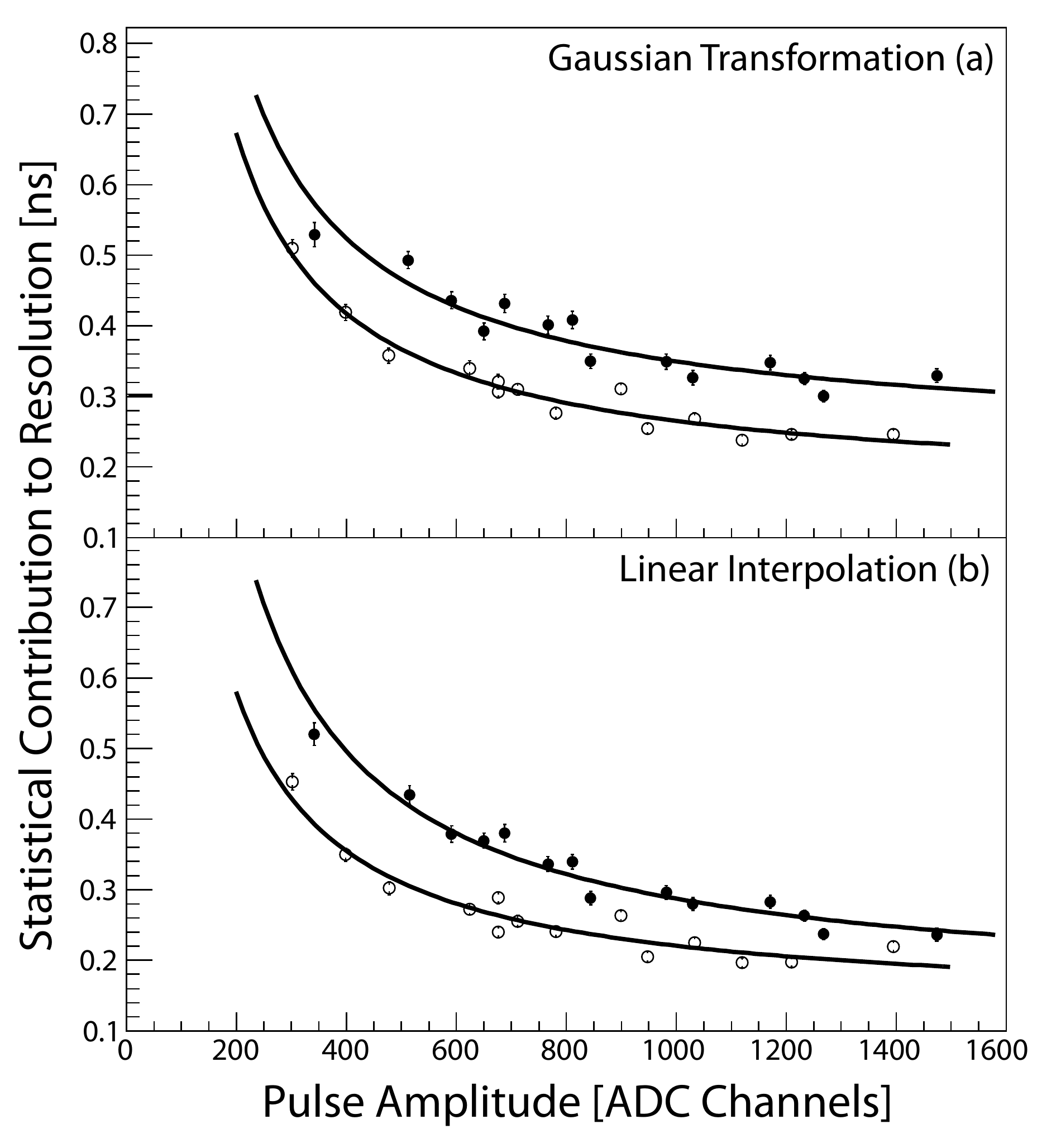}
\caption{The statistical contribution to the timing resolution of channels 1 (open circles) and 8 (filled circles) are 
plotted as a function of pulse amplitude for the Gaussian transformation (a) and linear interpolation (b) timing algorithms.  These two channels represent the highest and lowest resolution of all elements in the array -- measurements for the other six channels are between these extremes.
\label{fig:sigma_vs_pulse}
}
\end{figure}
\end{center}

Assuming the resolutions of the four control blocks are constant, the values of $\sigma_i$ for the four blocks 
with attenuated light input may be calculated using Eq.~\ref{eq:sigma_ij}.  The experiment is then repeated again with a different choice of control channels, allowing us to study all eight channels in the array.  The process is repeated while varying quantity of light incident on the channels under study in order to map out the dependence of the statistical contribution to the resolutions as a function of pulse amplitude.  The dependence is shown for the best and worst resolution channels in Fig~\ref{fig:sigma_vs_pulse} for the Gaussian transformation 
and linear interpolation method.  For relatively low pulse amplitudes the resolution is still better than 1~ns and appears to be in the range of 200-400~ps for larger pulses. 

The calculation of statistical contribution to the resolutions depends upon the choice of control channel used to determine $\sigma_{ij}$.  Any one channel may be compared to four different control channels. As a cross check on our procedure we varied the control channel used in our extraction of the statistical contributions to the resolution. This variation produced changes in measured $\sigma_i$ of less than 40~ps over the entire range of pulse amplitudes indicating that our procedure is robust.

The statistical contributions to the resolution were fit to
\begin{equation}
  \sigma(S_p) = \frac{a}{S_{p}} + b,
\label{eq:parametrization}
\end{equation}
where $a$ and $b$ are free parameters (as shown in Figure~\ref{fig:sigma_vs_pulse}). The average value for $a$ over all eight channels was determined to be $117 \pm 34$ ($114\pm46$)~channels$\cdot$ns while that for $b$ is 
$0.203 \pm 0.079$ ($0.155\pm0.077$)~ns for the Gaussian transformation (linear interpolation) algorithm.  At an amplitude of 500~mV the average statistical contribution to the resolution is $0.29\pm0.09$ ($0.24\pm0.08$)~ns, while at an amplitude of 100~mV the resolution is $0.62\pm0.16$ ($0.56\pm0.18$)~ns for the Gaussian transformation (linear interpolation) method.  The uncertainties of the results are derived from the standard deviation of the resolutions for all eight channels studied.  Both algorithms provide comparable resolution.

\begin{center}
\begin{figure}
\includegraphics[width=\linewidth]{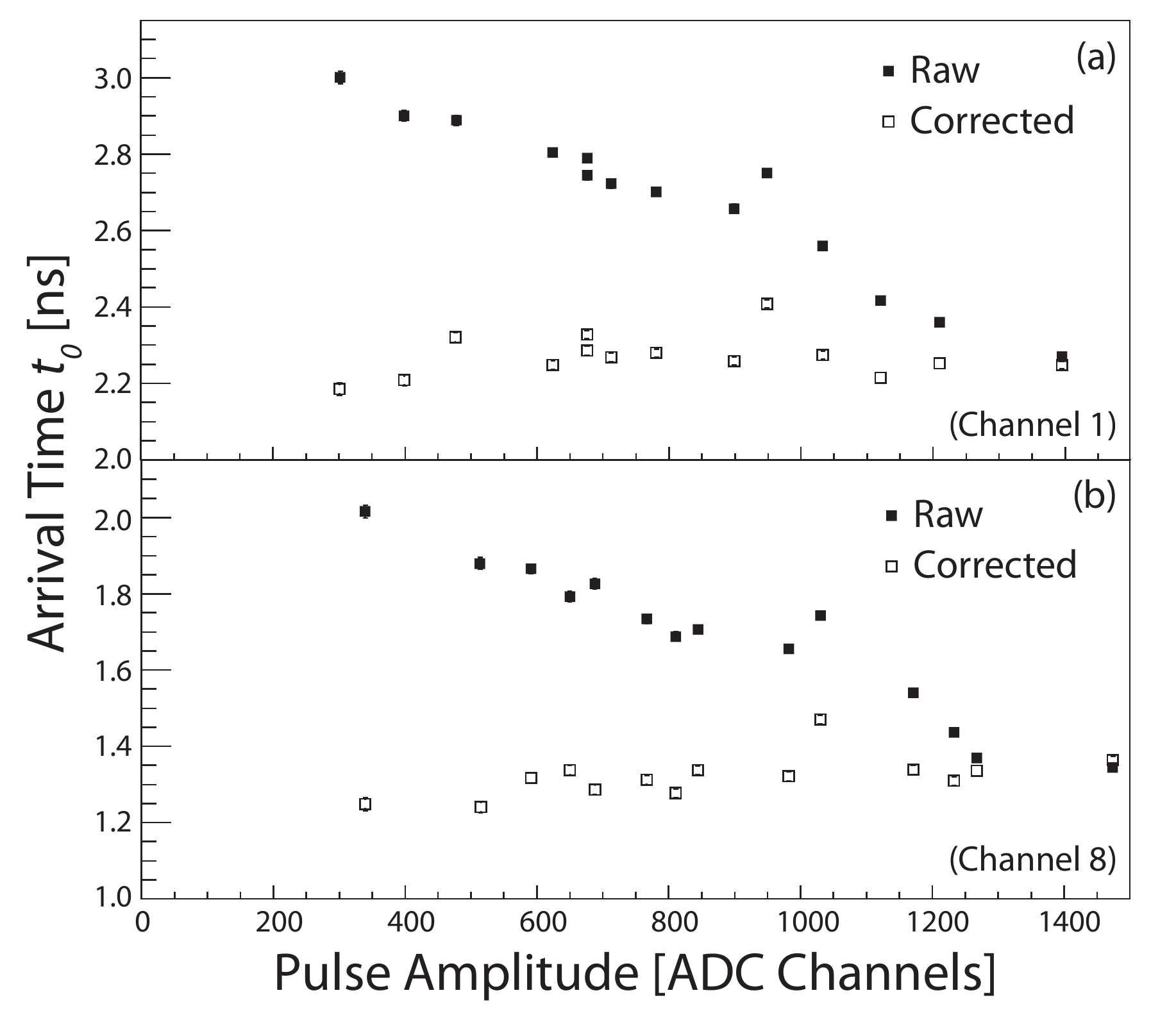}
\caption{The uncorrected $t_0$ values (filled squares) as a function of signal amplitude calculated from the linear interpolation method  for channels 1 (a) and 8 (b).  The values after the correction given by Eq.~\ref{eq:mu_correction} are shown as open squares. Other channels are consistent with these, and similar behavior is observed for the Gaussian transformation algorithm.
\label{fig:mean_vs_pulse}
}
\end{figure} 
\end{center}

The values of the average pulse arrival time $t_{0,i}$ calculated from the same data sets are shown in filled squares in
Fig.~\ref{fig:mean_vs_pulse}. These data show a significant systematic shift in $t_{0,i}$ as a function of pulse amplitude.  In order to account for this shift, which would otherwise cause instability in the measurement of $t_0$ under variations of the pulse amplitude, the values of $t_{0}$ for each channel were corrected by the procedure
\begin{equation}
  t_{0,i} \to t_{0,i} + \alpha S_p, 
\label{eq:mu_correction}
\end{equation}
where $\alpha$ is a single parameter, apparently characteristic of the algorithm and FEU~84-3 pulse shape. In this study, 
$\alpha$ was calculated to be about $-7 \times 10^{-4}$ ns/ADC counts, independent of channel and also digitization algorithm.  This correction, which is applied to the to the measured characteristic time, dramatically reduces the dependence of the characteristic time on pulse amplitude as shown in Fig.~\ref{fig:mean_vs_pulse} and, consequently, reduces the systematic contribution to the timing resolution.

The total contribution to the timing resolution from statistical and residual systematic fluctuations can be characterized as
\begin{equation}
  \sigma(S_{p}) = \sqrt{\left(\frac{a}{S_{p}} + b\right)^2 + c^2},
\label{eq:total_sigma}
\end{equation}
where $a$ and $b$ are determined as discussed above and $c$ is the error due to the residual systematic contribution to the resolution. By computing the standard deviation of the corrected $t_0$ for all channels over a range of pulse amplitudes (as is shown for two channels in Fig.~\ref{fig:mean_vs_pulse}) the value for $c$ was determined to be 0.05 ns, a minor contribution when added in quadrature of the statistical contributions given by $a$ and $b$.

\section{Conclusions and discussion}
\label{Summary}

We have presented two signal processing algorithms for converting flash ADC data into a measurement of the arrival time of an FEU~84-3 PMT pulse.  One algorithm utilizes the fact that the leading edge of the pulse can be described by a Gaussian, while the other uses a linear approximation to the leading edge in the vicinity of the time at which the pulse has reached half of its maximum amplitude.  We have measured the resolution of these algorithms and also the anticipated resolution with which we can measure a characteristic arrival time of a PMT pulse in the final GlueX detector configuration.

By devising a technique to delay and digitize the same pulse twice with a single-channel eight-bit FADC we studied the capabilities of the timing algorithms, independent of variations in pulse shape or transit time through the PMT.
At a pulse height around 1.0 V, 
corresponding to roughly 85\% of the full scale of an eight-bit FADC, we determine that
timing resolution of the Gaussian transformation algorithm is  in the range of 50-110~ps, while the resolution for the linear interpolation algorithm is 40-150~ps.  The variation in resolutions is attributable to variations between the true time of the signal peak and the phase of the sampling clock.  The time resolution that can be obtained by applying the two methods is only a fraction of the 4~ns time between samples.
 
The relevant timing resolution for a single channel in the GlueX calorimeter includes contributions from the timing algorithm, statistical contributions from pulse-to-pulse variations in shape and PMT transit time, and systematic contributions arising from dependence of measured arrival time on pulse height.  A nine-channel array was constructed to study these effects.  The statistical fluctuations in arrival time, which depend on pulse amplitude, dominate the timing resolution and were measured as a function of pulse amplitude for both algorithms.  The systematic shift in measured time as a function of pulse amplitude appears to be linear and consistent across all channels for both of the timing algorithms.  This allows for a simple, single-parameter correction to be performed, which reduces the systematic contribution to the timing resolution to about 50~ps, making it nearly negligible.  The linear interpolation algorithm performs slightly better than the Gaussian interpolation algorithm, although either would meet the resolution demands of the GlueX forward calorimeter.  Assuming the use of the linear interpolation algorithm, the estimated final timing resolution anticipated in the detector for FEU~84-3 PMT pulses with amplitudes of 100~mV (500~mV) is $0.57\pm 0.18$ ($0.24\pm 0.08$)~ns, where the central values include the quadrature-sum of statistical and systematic contributions to the resolution.

Since the algorithms rely on sampling the pulse on its leading edge and near the peak, we anticipate that measurements of the characteristic pulse time will remain robust in a high rate environment assuming PMT pulses do not overlap.  The digital signal processing electronics could further be modified to apply the algorithm to multiple detected peaks in a single flash ADC buffer.  The full FEU 84-3 pulse fits inside of a 40~ns window.  The 1-2\% of FCAL modules that are closest to the beam axis have expected background rates of $>1$~MHz.  For the highest rate modules the probability of pulse overlap may reach 10\%, resulting in some small background contamination.  For the remaining 98\% of the modules, the probability of overlap is expected to be negligible.

Both algorithms presented provide comparable timing resolution that is suitable for our application.  The difference in the two algorithms is in the implementation.  The Gaussian transformation algorithm relies on more complicated mathematical functions; however, has the advantage that it always uses the peak sample and the two samples immediately preceding the peak.  The linear interpolation algorithm, while mathematically more simple, requires a search and comparison of all samples on the leading edge to find the sample immediately before and after the point at which the signal level crosses half the maximum value.  The exact choice of algorithm for the GlueX application will depend on which technique can be most efficiently implemented in flash ADC FPGA.

\section{Acknowledgements}

We would like to acknowledge the members of the Jefferson Lab electronics and data acquisition group who have supported instrumentation used in this study:  D.~Abbott, F.~Barbosa, C.~Cuevas, H.~Dong, E.~Jastrzembski, B.~Raydo, and E.~Wolin.  P.~Smith of Indiana University both designed and provided support for the eight-bit FADC that was used in these studies.  This work was supported by the Department of Energy under contract DE-FG02-05ER41374.  Jefferson Science Associates, LLC operated Thomas Jefferson National Accelerator Facility for the United States Department of Energy under contract DE-AC05-06OR23177.


\begin{thebibliography}{9}

\bibitem{gluex} See, for example:  M.~R.~Shepherd,
 {\it Proceedings of the Tenth Conference on the Intersection of Particle and Nuclear Physics}, AIP Conf. Proc. 1182, 816 (2009); {\it http://www.gluex.org}.

\bibitem{Barbosa07}F. J. Barbosa {\it et al.},
% \textsl{A VME64x, 16-channel, Pipelined, 250MSPS Flash ADC with Switched Serial (VXS) Extension}, 
IEEE Nuclear Science Symposium, Hawaii, USA (2007).

\bibitem{Brunner}A. Brunner {\it et al.},
% \textsl{A Cockcroft-Walton base for the FEU84-3 photomultiplier tube},
Nucl. Instrum. Meth. {\bf A 414}, 466 (1998).

\bibitem{moment}
J. Albert {\it et al.}
% signal reconstruction for the MAGIC telescope. 
Nucl. Instrum. Meth. {\bf A 594}, 407 (2008); H.~Von Der Schmitt {\it et al.},
  %``PARALLEL PROCESSING OF FLASH ADC DATA FOR THE JET CHAMBER OF JADE,''
  Nucl.\ Instrum.\ Meth.\  {\bf A 263}, 206 (1988).

\bibitem{Teige_time}
S. Teige {\it et al.},
%\newblock {P}hotomultiplier pulse timing using flash analog to digital converters.
Internal memo, GlueX-doc 426 (2005).

\bibitem{cogan}
P. Cogan, {\it Proceedings of the 30th International Cosmic Ray Conference}, arXiv:0709.4208 [astro-ph] (2007).


\end{thebibliography}
\end{document}